\documentclass[letterpaper, 10 pt, conference]{ieeeconf}
\IEEEoverridecommandlockouts
\usepackage{graphicx}
\usepackage{amsmath}
\usepackage{amssymb}
\usepackage{amsfonts}
\usepackage{tabularx}
\usepackage[table]{xcolor}

\title{\LARGE \bf
Data Augmentation for Deep Learning-based Radio Modulation Classification
}

\author{Liang Huang$^{1}$, Weijian Pan$^{1}$, You Zhang$^{1}$, LiPing Qian$^{1}$, Nan Gao$^{2}$ and Yuan Wu$^{3}$	
\thanks{$^{1}$College of Information Engineering, Zhejiang University of Technology,
	{\tt\small \{lianghuang, 2111803213, lpqian\}@zjut.edu.cn, yzhang\_zjut@163.com}.
}%
\thanks{$^{2}$College of Computer Science and Technology, Zhejiang University of Technology,
	{\tt\small \{gaonan\}@zjut.edu.cn}, (Corresponding Author).
}%
\thanks{$^{3}$State Key Laboratory of Internet of Things for Smart City, Department of Computer Information Science, University of Macau,
		{\tt\small \{ywucisum\}@gmail.com}.
}
}

\begin{document}

\maketitle

\thispagestyle{empty}
\pagestyle{empty}

\begin{abstract}
Deep learning has recently been applied to automatically classify the modulation categories of received radio signals without manual experience. However, training deep learning models requires massive volume of data. An insufficient training data will cause serious overfitting problem and degrade the classification accuracy. To cope with small dataset, data augmentation has been widely used in image processing to expand the dataset and improve the robustness of deep learning models. However, in wireless communication areas, the effect of different data augmentation methods on radio modulation classification has not been studied yet. In this paper, we evaluate different data augmentation methods via a state-of-the-art deep learning-based modulation classifier. Based on the characteristics of modulated signals, three augmentation methods are considered, i.e., rotation, flip, and Gaussian noise, which can be applied in both training phase and inference phase of the deep learning algorithm. Numerical results show that all three augmentation methods can improve the classification accuracy. Among which, the rotation augmentation method outperforms the flip method, both of which achieve higher classification accuracy than the Gaussian noise method. Given only 12.5\% of training dataset, a joint rotation and flip augmentation policy can achieve even higher classification accuracy than the baseline with initial 100\% training dataset without augmentation. Furthermore, with data augmentation, radio modulation categories can be successfully classified using shorter radio samples, leading to a simplified deep learning model and shorter the classification response time. 
\end{abstract}

\section{Introduction}
\label{sec:intro}
\PARstart{B}{enefiting} from the improvement of computing power and big data, deep learning has achieved unprecedented development in many applications, i.e., speech and audio processing \cite{wang2018supervised}, natural language processing \cite{7243232}, object detection \cite{li2016deepsaliency}, and so on. In recent years, it also achieves dramatic development in the field of wireless communications, e.g., modulation classification \cite{ali2017automatic}, symbol detection \cite{Ye2017Power}, end-to-end communication \cite{anderson2018deepmod}, and mobile edge computing \cite{huang2019deep}, \cite{huang2018distributed}, \cite{huang2019multi}.

Deep learning-based modulation classification automatically and efficiently classify received signals without prior knowledge. Modulation classification is a fundamental step for many applications in wireless communication systems, such as spectrum management in cognitive communication systems \cite{ali2016advances} and unauthorized signal detection in secure communications \cite{dobre2007survey}, \cite{jang2018performance}. Traditional modulation classification method either requires high computational complexity or greatly depends on manual operations \cite{dobre2007survey}. Recently, deep learning is successfully introduced to classify signals \cite{triantaris2019automatic}, \cite{o2018over}, \cite{peng2018modulation}, \cite{tang2018digital}, \cite{ramjee2019fast}, \cite{rajendran2018deep}, which feeds raw signal data or its transforms into a deep neural network and instantly obtains the modulation category at the network output. It achieves higher classification accuracy than traditional methods for automatic modulation classification based on expert features such as higher order cumulants based features \cite{o2016convolutional}, while requiring a little extra computational overhead computation time.

Although deep learning-based approaches can greatly improve the performance of the modulation classifier, it requires a large volume of training radio samples. However, in practice, collecting a large amount of high quality and reliable training radio samples sometimes is costly and difficult. Data augmentation has been widely used to deal with lack of training data by artificially expanding the training dataset with label preserving transformation. Different data augmentation methods have been proposed in the literature, i.e., random cropping, rotation and mirroring in image classification \cite{krizhevsky2012imagenet}, \cite{he2016deep} and pitch shifting, time stretching and random frequency filtering in speech recognition\cite{salamon2017deep}. For deep learning-based radio modulation classification, data augmentation can improve its invariant, especially for small radio signal dataset.

\begin{figure}[t!] 
	\centering
	\includegraphics[width=3.3 in]{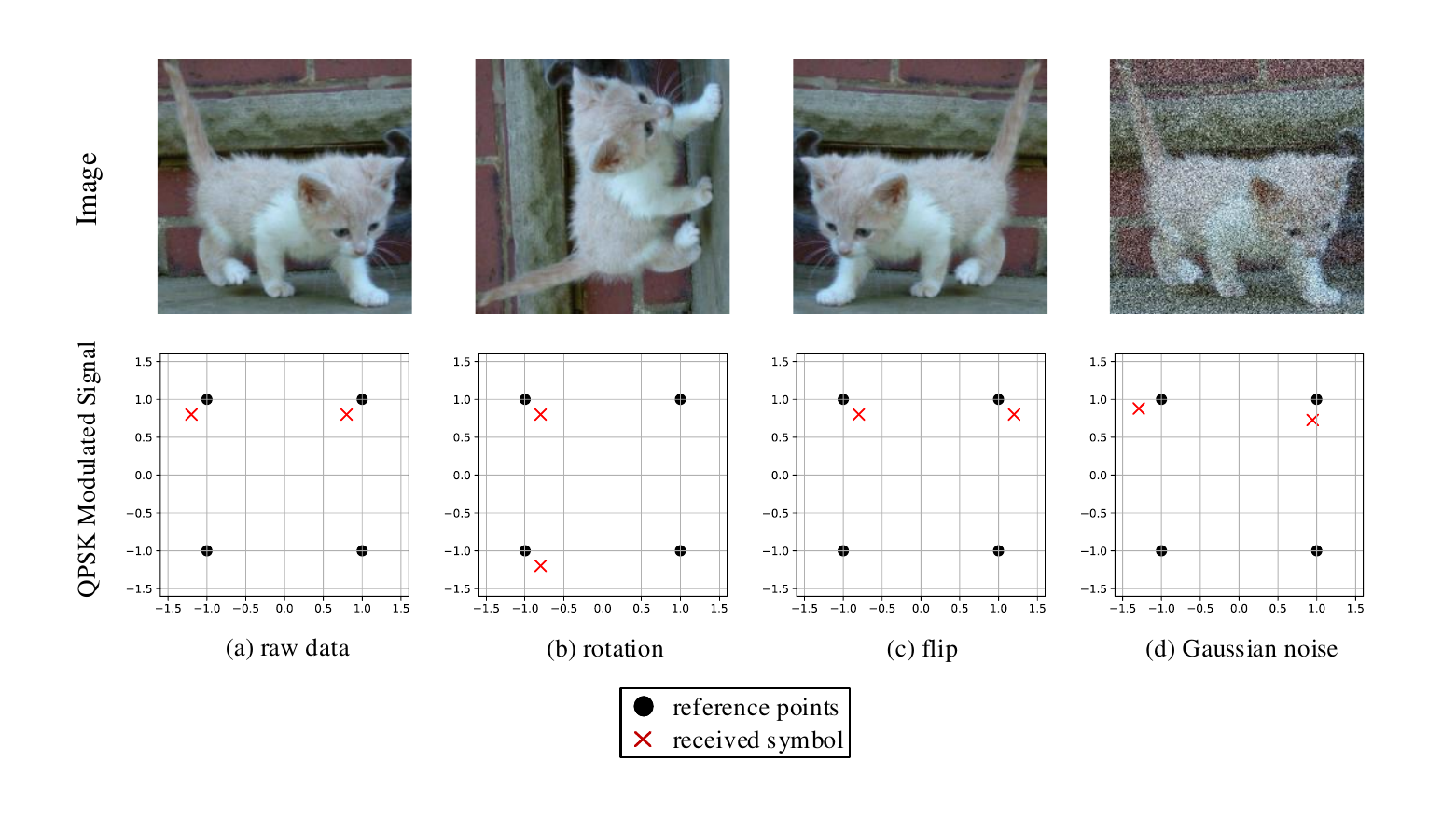}
	\caption{Different data augmentation methods for both image and modulated signal: (a) raw data, (b) rotation, (c) flip, (d) Gaussian noise.} \label{fig1}
\end{figure}

Augmenting modulated radio signal is similar to augment images as shown in Fig.~\ref{fig1}. Specifically, we consider three basic augmentation methods, i.e. rotation, flip, and Gaussian noise, for both an image and a quadrature phase-shift keying (QPSK) modulated radio signal sample illustrated in constellation diagram. For the image, after rotation or flip augmentation, the same cat is displayed but from different viewpoints. In the constellation diagram of the QPSK modulated radio signal, the black circles indicate four ideal reference points, and the red crosses are the received symbols which are shifted due to the imperfection of transmitter/receiver hardware and wireless channel \cite{sankhe2019oracle}. In Fig.~\ref{fig1}, we consider two received symbols with positive phase shift (1, 1) and (-1, 1), which are counter-clockwise shifted from their reference points. In wireless communication, each received symbol will be demodulated and mapped to one of the reference points based on the transmitted content. After rotation augmentation, two new symbols (-1, 1) and (-1, -1) are generated as shown Fig.~\ref{fig1}(b), which are also positively phase shifted. Therefore, for the radio modulation classification task considered in this paper, rotating the modulated radio signal is similar to rotating an image, without losing features for classification. However, flipping the radio signal generates two new QPSK modulated symbols whose phases are negatively shifted in the clockwise direction, as shown in Fig.~\ref{fig1}(c). Although both rotation and flip augmentation methods achieve similar accuracy improvements for image classification \cite{taylor2017improving}, \cite{shijie2017research}, it is an open question about which one is preferred for radio modulation classification. After the Gaussian noise augmentation, the image is full of 'snow' and the received radio symbols are deviated as shown in Fig.~\ref{fig1}(d). Can all these three augmentation methods improve the classification accuracy for deep learning-based radio modulation classification? To the best of our knowledge, the effect of different data augmentation methods on radio modulation classification has not been evaluated yet.

In this paper, we study data augmentation methods for deep learning-based radio modulation classification. Specifically, a state-of-the-art deep learning-based modulation classifier, is used to automatically classify the modulation category of each radio signal sample. Based on the characteristics of the modulated signal, we study three augmentation methods, i.e., rotation, flip, and Gaussian noise. After extensive numerical evaluations on an open radio signal dataset, we obtain the following contributions:

(1)	We propose algorithms to augment radio signals at both training phase and inference phase of the deep learning algorithm, which achieves around 2.5\% improvement on the baseline in terms of classification accuracy.

(2)	We discover that the rotation augmentation method outperforms the flip method, both of which achieve higher classification accuracy than the Gaussian noise method.

(3)	We propose a joint augmentation policy with both rotation and flip methods for insufficient training dataset. Given only 12.5\% of training dataset, the joint augmentation method expands the dataset to be a size of 75\% of the initial dataset and achieves an even higher classification accuracy than the baseline with 100\% training dataset without augmentation.

(4) With data augmentation, we successfully classify radio samples by using only one half of the sampling points. Therefore, the deep learning model can be simplified with a significantly reduced inference complexity. Furthermore, in the future field deployment, the modulation category can be successfully classified upon receiving only half number of radio sampling points, which greatly reduces the classification response time.

The remainder of this paper is organized as follows. Section II presents related work. Section III provides an overview of the studied radio signal dataset and the deep learning-based modulation classifier. We introduce three data augmentation methods in Section IV and propose an algorithm to augment signals at both deep learning phases in Section V. In Section VI, we present the simulation setup and the final experimental results. We finally conclude this paper in Section VII. 

\section{Related Work}
\subsection{Deep Learning in Radio Modulation Classification}
Deep learning has been applied to automatically classify radio modulation categories in recent literature. By converting radio signals into images, two convolutional neural network (CNN)-based deep learning models, GoogleNet \cite{szegedy2015going} and AlexNet \cite{krizhevsky2012imagenet}, originally developed for image classification, are used for modulation classification \cite{peng2018modulation}, \cite{tang2018digital}. The modulation classification accuracy is further improved by a modified deep residual network (ResNet) \cite{o2018over}, which is fed with the modulated in-phase (I) and quadrature phase (Q) signals. Considering channel interference, the CNN structure also achieves a considerable classification accuracy \cite{triantaris2019automatic}. In addition to the CNN-based models, the Long Short-Term Memory (LSTM) architecture with time-dependent amplitude and phase information can achieve the state-of-the-art classification accuracy \cite{rajendran2018deep}. To reduce the training time of deep learning models, different subsampling techniques are investigated in \cite{ramjee2019fast} which reduce the dimensions of the input signals.

\subsection{Data Augmentation in Deep Learning}
Data augmentation is widely used in deep learning algorithms to increase the diversity of training dataset, prevent model overfitting, and improve the robustness of the model. For image classification tasks, generic data augmentation methods include flip, rotation, cropping, color jittering, edge enhancement, and Fancy PCA \cite{taylor2017improving}. Other complex data augmentation methods synthesize a new image from two training images \cite{inoue2018data} or from Generative Adversarial Nets (GAN) \cite{antoniou2017data}. Although there are many augmentation methods for images, AutoAugment \cite{cubuk2018autoaugment} is proposed to automatically search for augmentation policies based on the dataset. In addition to images, augmentation methods such as synonym replacement, random insertion, random swap, and random deletion are used for text classification \cite{wei2019eda}, where the same accuracy as normal in all training data is achieved when only half of the training data is available. For speech recognition tasks, training audio is augmented by changing the audio speed \cite{ko2015audio}, warping features, masking blocks of frequency channels, and masking blocks of time steps \cite{park2019specaugment}.

There are few related works on data augmentation for radio modulation classification in the literature. The most related work is a GAN based data augmentation method proposed in \cite{tang2018digital}. The authors first converted the signal samples into Contour Stellar Images which were further used to train the GAN network so as to generate new signal training samples. With GAN-based augmentation, the modulation classification accuracy is improved by no more than 6\%. However, training GAN network still requires sufficient signal samples to guarantee the convergence. Moreover, as reported in \cite{tang2018digital}, the classification accuracy based on augmented dataset is lower than on real dataset with the same amount of signal samples. Therefore, an efficient augmentation method for insufficient radio signal dataset is still absent. 

\section{Preliminaries}

\begin{figure*}[t!] 
	\centering
	\includegraphics[width=7.16 in]{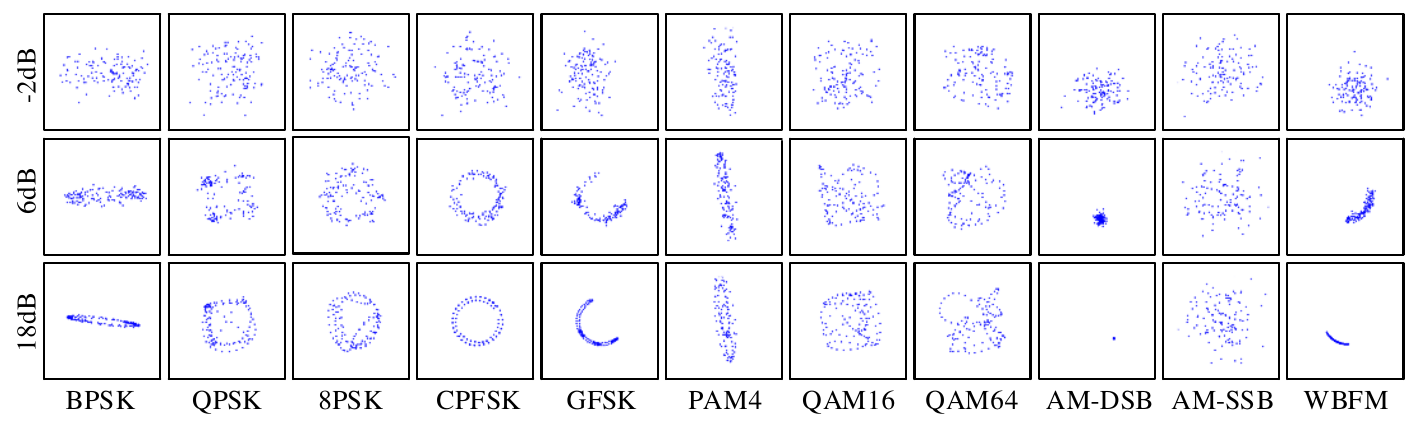}
	\caption{Constellation diagrams of 11 modulated signals \cite{o2016radio} under different SNRs.}\label{fig2}
\end{figure*}

In this section, we introduce the radio signal dataset and the architecture of the state-of-the-art LSTM model \cite{hochreiter1997long}, which will be used to evaluate different data augmentation methods presented in Sec. IV.

\subsection{Radio Signal Dataset}
We evaluate the radio signal modulation classification based on an open radio signal dataset, RadioML2016.10a \cite{o2016radio}. The radio signals in the dataset consider sample rate offset, center frequency offset, multi-path fading and additive white Gaussian noise. Specifically, there are 220,000 modulated radio signal segments belonging to 11 different modulation categories, i.e., binary phase-shift keying (BPSK), QPSK, eight phase-shift keying (8PSK), continuous phase frequency-shift keying (CPFSK), Gauss frequency-shift keying (GFSK), pulse-amplitude modulation four (PAM4), quadrature amplitude modulation 16 (QAM16), quadrature amplitude modulation 64 (QAM64), double-sideband AM (AM-DSB), single-sideband AM (AM-SSB) and wideband FM (WB-FM). Each radio signal sample is composed of 128 consecutive modulated in-phase (I) signal and quadrature phase (Q) signal. The labels of each signal sample include its value of signal-to-noise ratio (SNR) and its corresponding modulation category. There are total 20 different SNRs ranging from -20dB to 18dB with a step size of 2dB. In the dataset, these 220,000 signal samples are uniformly distributed among 11 modulation categories and 20 SNRs. In other words, there are 1,000 signal samples for each modulation category at each SNR. In Fig.~\ref{fig2}, we plot examples of 11 modulation categories in forms of constellation diagrams under different SNRs. In the following subsection, we introduce a deep learning algorithm which automatically predicts the radio's modulation category based on its raw I/Q signals.

\subsection{LSTM Network Architecture}
\begin{figure}[t!] 
	\centering
	\includegraphics[width=3.3 in]{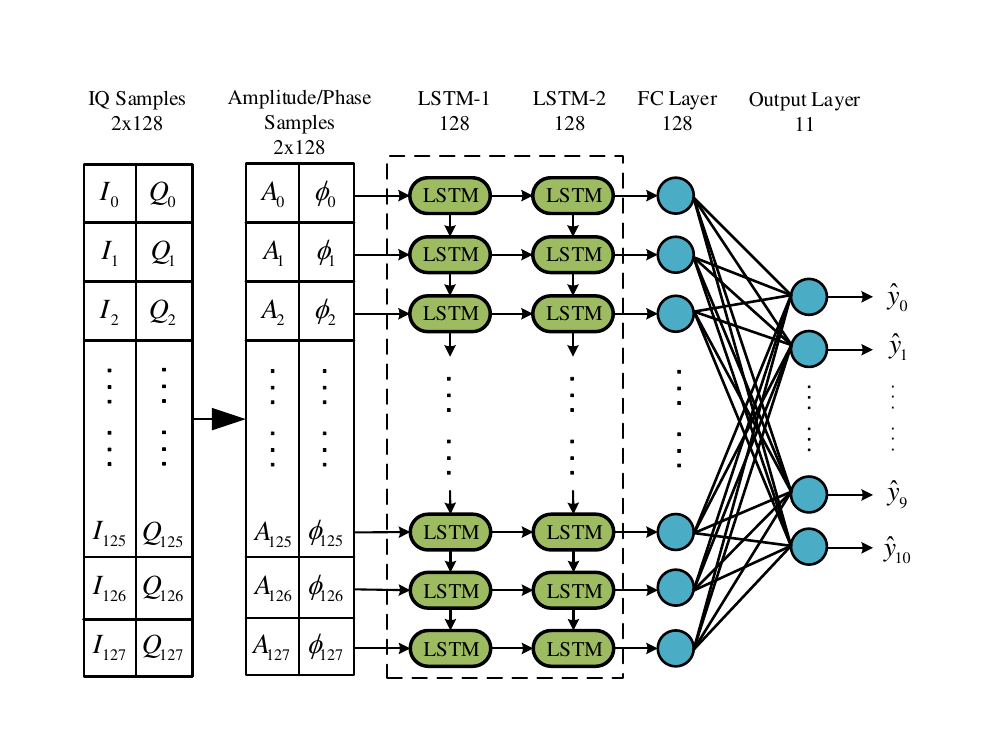}
	\caption{The architecture of LSTM network.} \label{fig3}
\end{figure}

LSTM is a special category of Recurrent neural network (RNN), which is widely used to process time series data. Benefited from a specific LSTM memory cell mechanism, LSTM effectively solves the exploding and vanishing gradient problem of traditional RNN during training process and learns long-term dependencies in sequential data. The LSTM memory cell mainly consists of a forget gate, an input gate and a update gate \cite{gers1999learning}, which implement selective retention and discard of input information.

The LSTM network takes each data sample with consecutive modulated in-phase (I) and quadrature phase (Q) signals as input and maps them to a specific modulation category, whose architecture is shown in Fig.~\ref{fig3}. Specifically, the modulated I/Q signals are first converted into amplitudes and phases \cite{rajendran2018deep}, as:
\[\left\{ \begin{array}{l}{\rm A}{\rm{ = }}\sqrt {{I^2} + {Q^2}} \\\phi {\rm{ = arc}}\tan ({\rm{Q}}/I)\end{array} \right.,\]
where $A$ and $\phi $ represent the amplitude and phase of the modulated signal, respectively. The obtained signals are then fed into a two-layer LSTM network to extract characteristic features, where each layer has 128 LSTM cells. Finally, a fully connected layer with Softmax function is used to map the radio signal sample to one of these 11 modulation categories. Adam optimizer \cite{kingma2014adam} with dynamic learning rate is used to minimize the cross-entropy loss as follows:
\[\ell  =  - \sum\limits_{k = 1}^K {{y_k}\log ({{\hat y}_k})},\]
where $K$ is the number of classes, ${y_k}$ represents the ground truth label, and ${\hat y_k}$  denotes the probability that the input sample will be predicted as $k{\rm{ - th}}$ class.

\section{Data Augmentation Methods}
\label{sec:guidelines}
Data augmentation is a method widely used in deep learning because it improves the generalization ability of the model and alleviates overfitting. In this section, we describe in detail three data augmentation methods for modulation signal recognition, including rotation, flip, and Gaussian noise. The dataset is expanded by a scale factor $N$.

\begin{figure}[t!] 
	\centering
	\includegraphics[width=3.3 in]{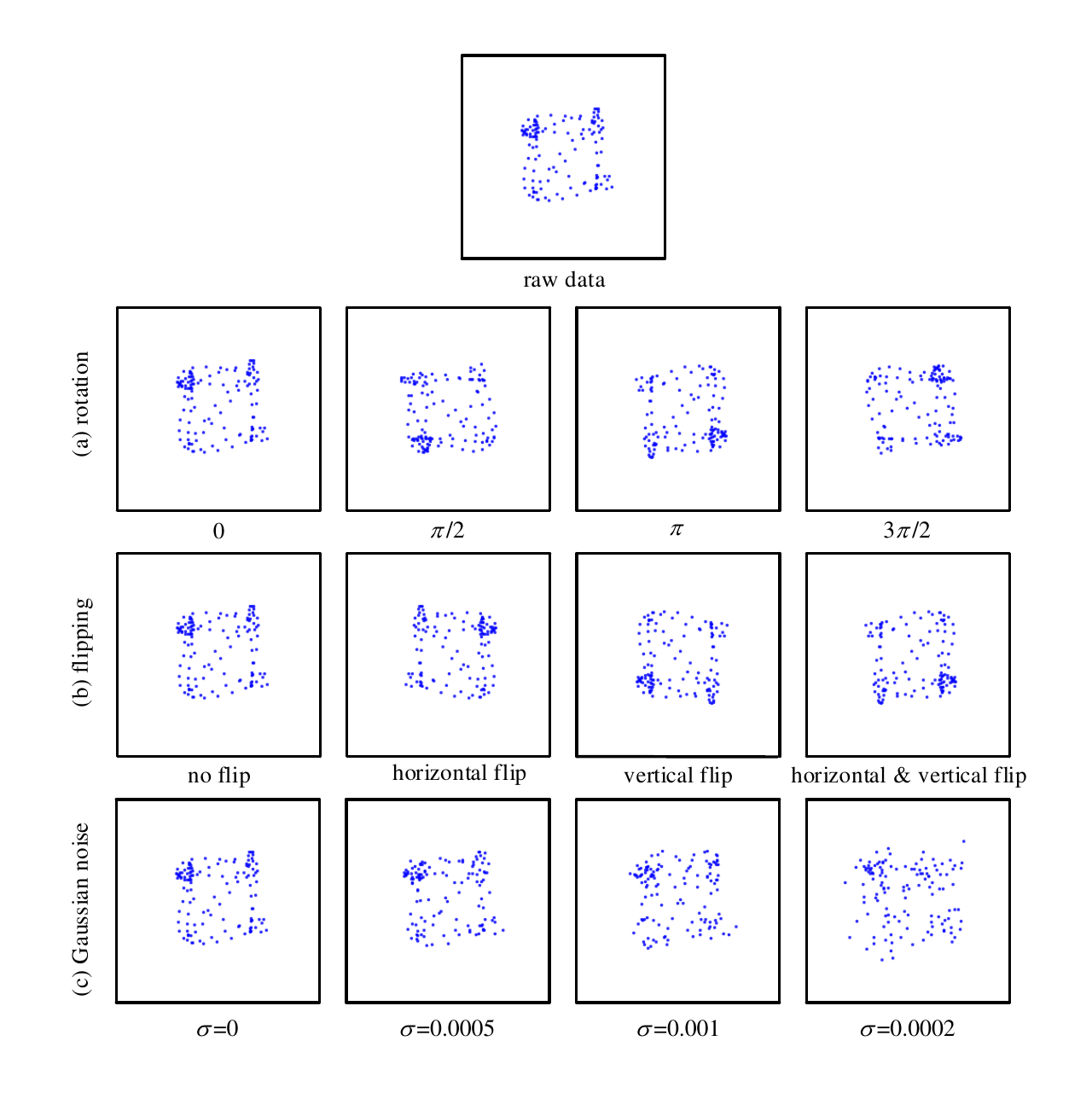}
	\caption{Constellation diagram of an QPSK radio signal sample with different data augmentation methods.} \label{fig4}
\end{figure}

\subsection{Rotation}
By rotating a modulated radio signal $(I,Q)$ around its origin, we obtain augmented signal sample $(I',Q')$ as follows:
\[\left[ \begin{array}{l}
I'\\
Q'
\end{array} \right] = \left[ \begin{array}{l}
\cos \theta \ \ {\rm{  - sin}}\theta \\
\sin \theta \ \ \ \ \ {\rm{  cos}}\theta 
\end{array} \right]\left[ \begin{array}{l}
I\\
Q
\end{array} \right],\]
where $\theta$ is the angle of rotation. In this paper, the radio signal is rotated in the counter-clockwise direction by $0$, $\pi {\rm{/2}}$, $\pi$, and ${\rm{3}}\pi {\rm{/2}}$. In Fig.~\ref{fig4}(a), we plot the constellation diagram of an QPSK sample where one set of raw data is augmented into four radio signal samples.

\subsection{Flip}
For a given modulated radio signal $(I,Q)$ , we define the horizontal flip by switching the $I$ value to its opposite, as:
\[\left[ \begin{array}{l}
I'\\
Q'
\end{array} \right] = \left[ \begin{array}{l}
- I\\
\ \ Q
\end{array} \right],\]
and define the vertical flip by switching the $Q$ value to its opposite, as:
\[\left[ \begin{array}{l}
I'\\
Q'
\end{array} \right] = \left[ \begin{array}{l}
\ \ \ I\\
- Q
\end{array} \right],\]
to augment the radio signals. We can perform horizontal flip, vertical flip, or both flips at the same time such that the signal dataset is expanded by a scale factor $N = 4$, as shown in Fig.~\ref{fig4}(b).

\subsection{Gaussian Noise}
By adding a Gaussian noise ${\cal N}(0,{\sigma ^2})$ to the modulated radio signal $(I,Q)$, we obtain the augmented signal sample $(I',Q')$ as:
\[
\left[ \begin{array}{l}
I'\\
Q'
\end{array} \right] = \left[ \begin{array}{l}
I\\
Q
\end{array} \right] + {\cal N}(0,{\sigma ^2}),
\]
where ${\sigma ^2}$ is the variance of noise. In Fig.~\ref{fig4}(c), we show the augmented signal samples by adding Gaussian noise with different standard deviations $\sigma {\rm{ = 0}}$, $\sigma {\rm{ = 0.0005}}$ $\sigma {\rm{ = 0.001}}$ and $\sigma {\rm{ = 0.002}}.$ For each data augmentation method, the original radio signal dataset is expanded by a default scale factor $ N = 4$, as illustrated in Fig.~\ref{fig4}. Note that the Gaussian noise data augmentation is supposed to significantly expand the dataset by choosing enough different values of $\sigma$. However, in the next section, we show that the Gaussian noise data augmentation is not preferred for radio data augmentation.

\section{Data Augmentation Time}
The execution of a deep learning algorithm includes training phase and inference phase. Data augmentation can be performed in both phases, resulting in three possible combinations of augmentations, i.e., test-time augmentation, train-time augmentation, and train-test-time augmentation.

\subsection{Train-time augmentation}
Train-time augmentation performs data augmentation during the training stage of the model. That is the training dataset is augmented and expanded by a scale factor $N$ while the test dataset remains the same. Taking the rotation data augmentation as an example, the training dataset is expanded from 110,000 radio signal samples to 440,000 samples after train-time augmentation. In general, a larger size of training dataset leads to a higher modulation classification accuracy.

\subsection{Test-time augmentation}
Test-time augmentation fuses features of all augmented radio signal samples in inference phase. In the inference phase, one radio signal sample $(I,Q)$ in the test dataset is augmented into $N$ samples $\{ {(I',Q')^n}|n \in N\}$. Then each augmented sample ${(I',Q')^n}$ is fed into the LSTM network, and we obtain a vector of corresponding predicted probabilities $\hat y_k^n$. The predicted modulation category is decided through summing the predicted probabilities $\hat{y}_k^n$ over all $N$ augmented samples and choosing the one with maximum conference \cite{zheng2019fusion}, as:
\[\mathop {\arg \max }\limits_{1 \le k \le K} \sum\limits_{n = 1}^N {\hat{y}_k^n}.\]

\subsection{Train-test-time augmentation}
Train-test-time augmentation conducts both train-time augmentation and test-time augmentation, where both training and test datasets are augmented and expanded by a factor $N$.

\begin{figure}[t!] 
	\centering
	\includegraphics[width=3.3 in]{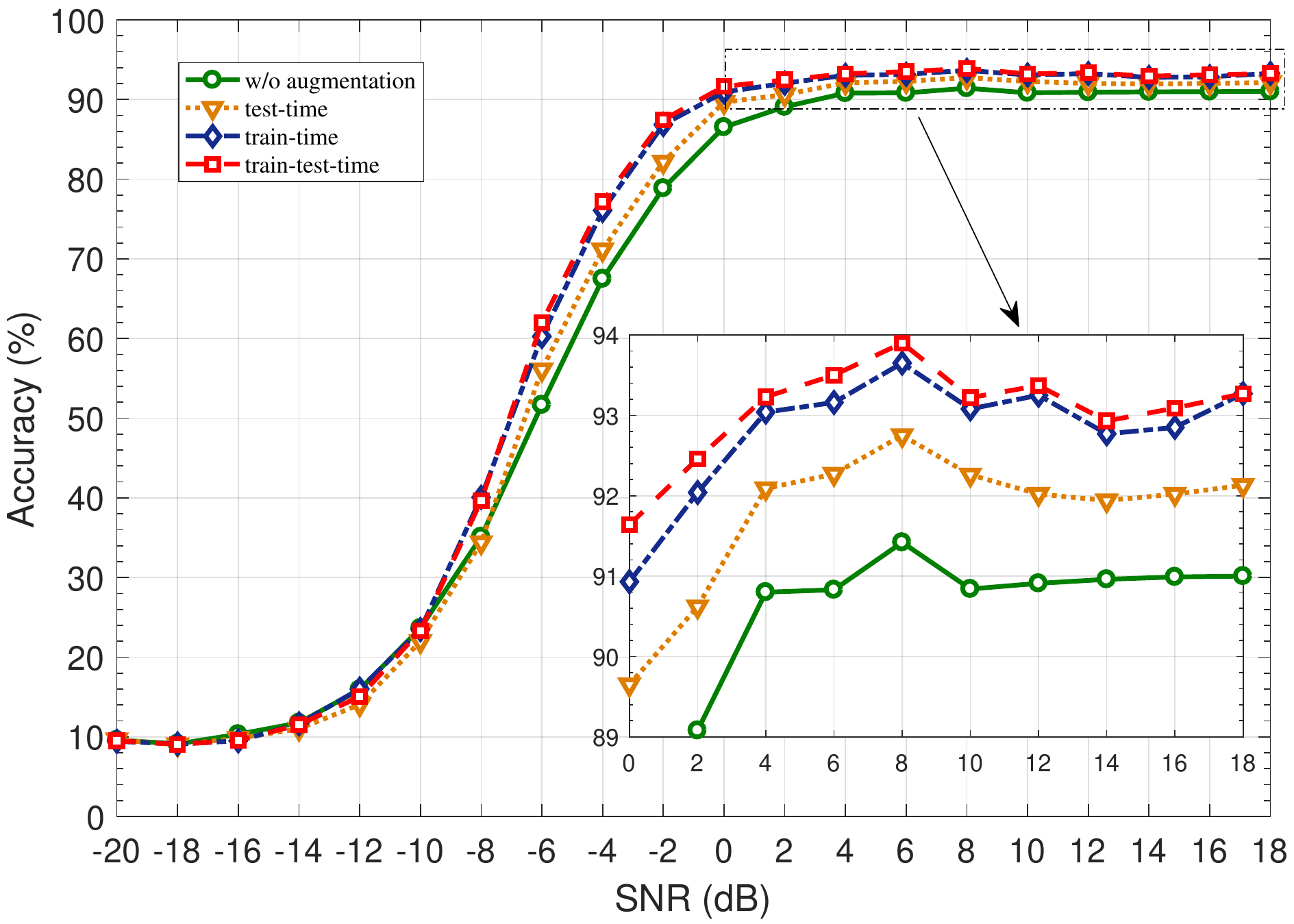}
	\caption{Classification accuracy under different augmentation times.} \label{fig5}
\end{figure}

In Fig.~\ref{fig5}, we numerically study the performance of the data augmentation at different phases, where the rotation augmentation with a scale factor $N=4$ is considered. Comparing with the baseline without augmentation, augmentations at different phases all improve the classification accuracy when the SNR is greater than -10 dB. The train-time augmentation achieves better performance than test-time augmentation, and the train-test-time augmentation generates the highest accuracy. Specifically, comparing with the baseline, the train-test-time augmentation improves the modulation classification accuracy by 8.87\% when SNR is -6dB and by about 2.2\% when SNR is greater than 4 dB. In the following numerical studies, we use the train-test-time augmentation by default.

\section{Augmentation Performance}
In this section, we numerically study the performance of different radio data augmentation methods in terms of modulation classification accuracy. The open dataset, RadioML2016.10a, is divided equally into a training dataset and a test dataset, each containing 110,000 radio signal samples. In order to avoid overfitting, we set dropout rate to be 0.5 at both two LSTM layers. The number of training epoch is 80 and the mini-batch size is 128. The value of the learning rate is initially set as 0.001 and is halved when the training accuracy is not improved during three consecutive epochs. The model is implemented based on PyTorch \cite{paszke2017automatic}.

\subsection{Augmentations On Full Dataset}

\begin{figure}[t!] 
	\centering
	\includegraphics[width=3.3 in]{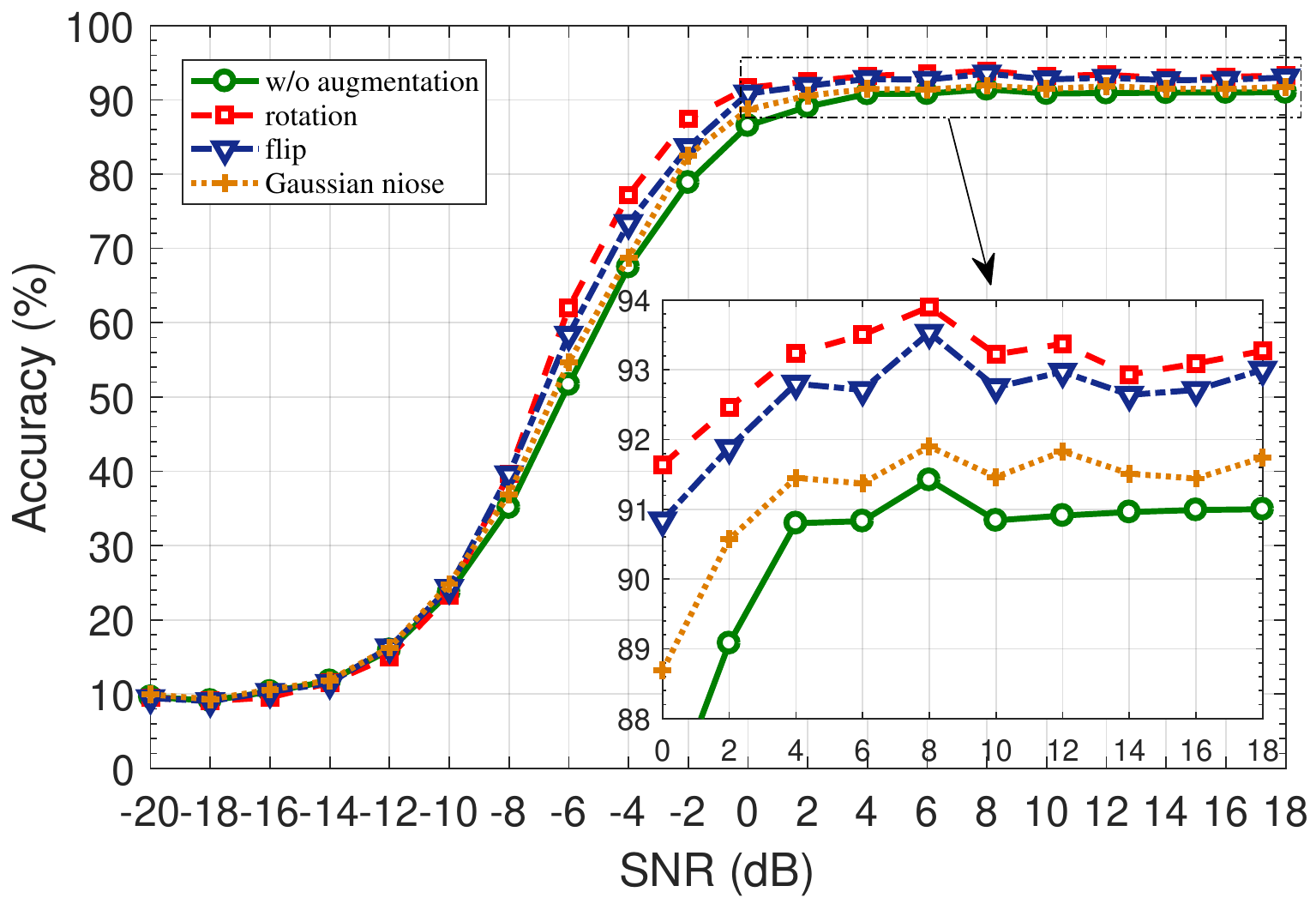}
	\caption{Classification accuracy under different augmentation methods.} \label{fig6}
\end{figure}

\begin{figure*}[t!] 
	\centering
	\includegraphics[width=7. in]{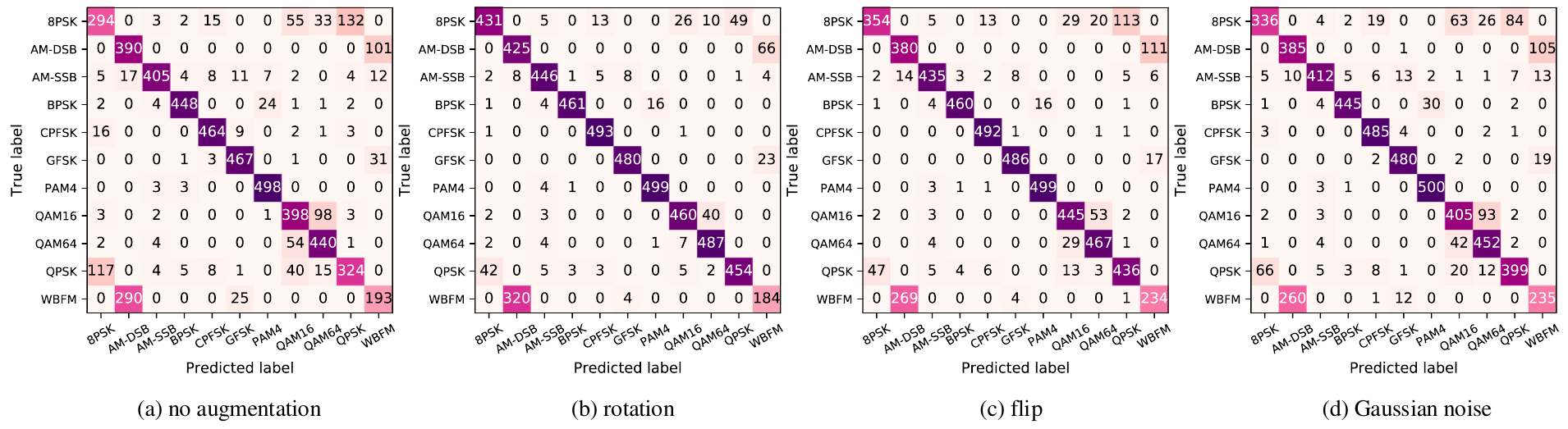}
	\caption{Confusion matrices under different data augmentation methods with 100\% training dataset when SNR is -2dB.}\label{fig7}
\end{figure*}

\begin{figure*}[t!] 
	\centering
	\includegraphics[width=7. in]{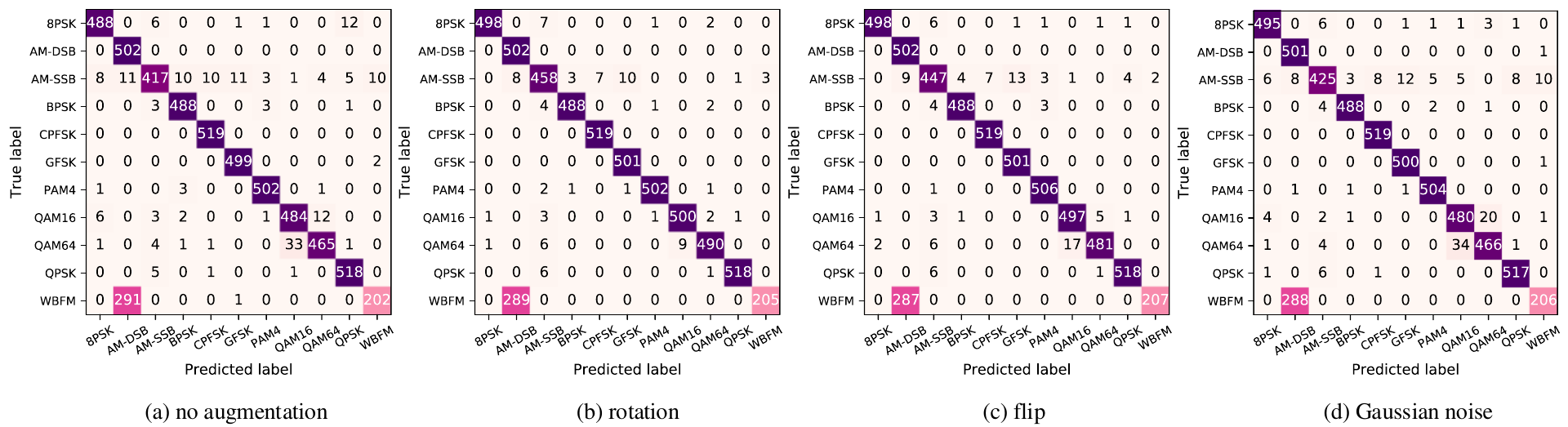}
	\caption{Confusion matrices under different data augmentation methods with 100\% training dataset when SNR is 18dB.}\label{fig8}
\end{figure*}

In Fig.~\ref{fig6}, we study the modulation classification accuracies of the LSTM model after deploying all three data augmentation methods presented in Sec. VI. Comparing with the baseline without augmentation, all augmentation methods improve the classification accuracy when the SNR is greater than -10dB, especially for the rotation data augmentation and flip data augmentation. In particular, the rotation data augmentation method achieves the greatest improvement by 8\% when SNR is between -6dB and -2dB and by about 2\% at higher SNR ($\ge$4dB). Meanwhile, the Gaussian noise data augmentation performs better at lower SNR when it is between -16dB and -10dB. Intuitively, adding Gaussian noise reduces the SNR of the original data sample which in turn generates more signal samples with low SNR. However, the improvement is trivial since the resulting classification accuracy is too small, less than 2\% when SNR is smaller than 10 dB. Therefore, rotation data augmentation and flip data augmentation are more preferred for radio signals in modulation classification.

To further evaluate the improvements of different augmentation methods on classification accuracy, we present the corresponding confusion matrices these at low SNR (-2dB) and high SNR (18dB) in Fig.~\ref{fig7} and Fig.~\ref{fig8}, respectively. Most values at diagonal entries of these matrices are increased after argumentation, which means the modulation classification accuracy are improved. Specifically, the proposed augmentation methods successfully reduce the confusion between QAM16 and QAM64 and solve the short-time observation problem presented in \cite{o2017introduction}. At low SNR, the LSTM model is difficult to classify 8PSK and QPSK, whose classification accuracy is greatly improved after rotation augmentation as shown in Fig.~\ref{fig7}. At high SNR, the accuracy of the LSTM model is mainly limited by the confusion between AM-DSB and WBFM, which dues to frequent radio samples without information in the dataset \cite{o2017introduction}. In general, rotation and flip achieve better classification accuracy than Gaussian noise for all modulation categories.

\subsection{Augmentations On Partial Dataset}
\begin{figure}[t!] 
	\centering
	\includegraphics[width=3.3 in]{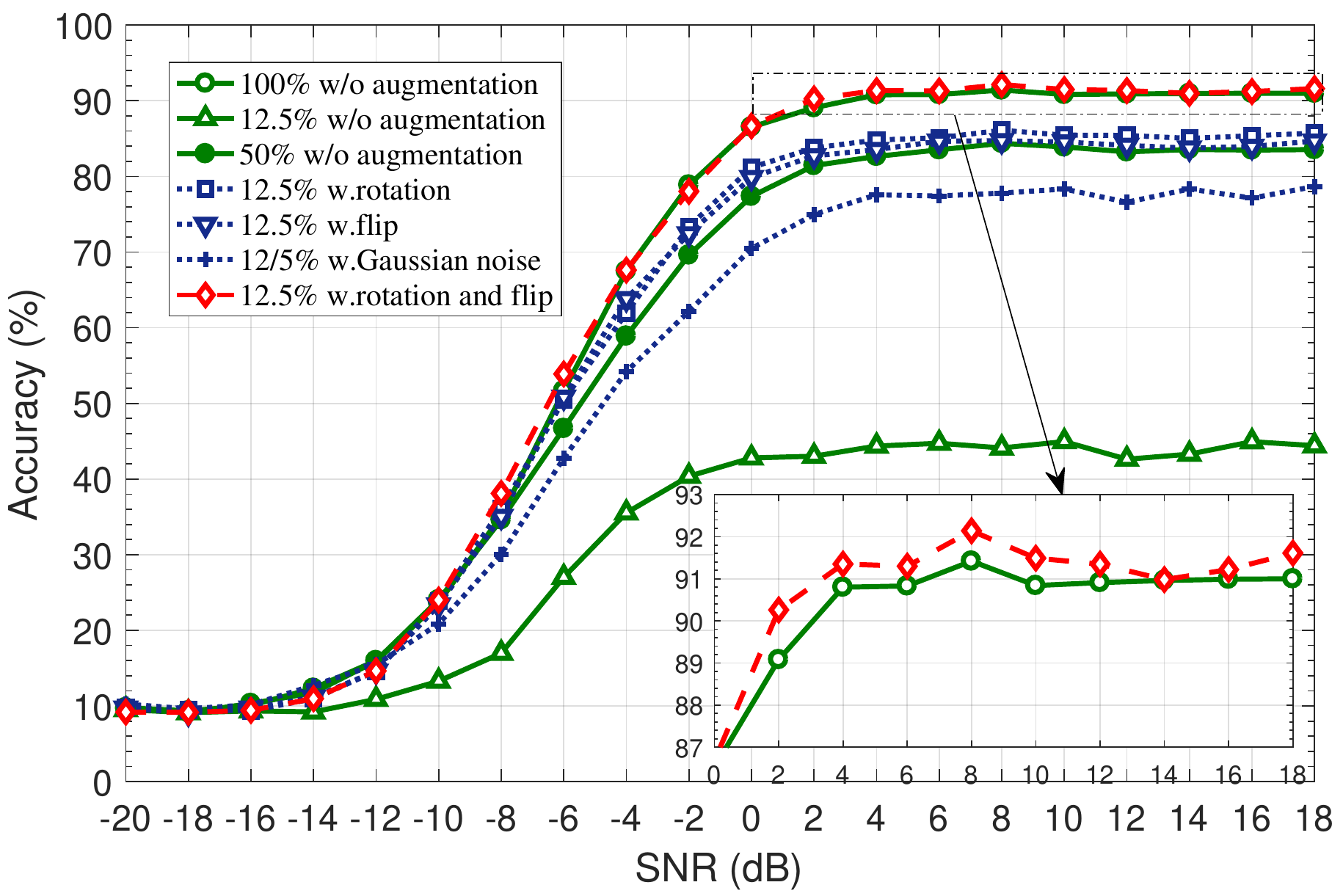}
	\caption{Classification accuracy versus different data augmentation methods under different SNRs with 12.5\% training dataset.} \label{fig9}
\end{figure}

In Fig.~\ref{fig9}, we further study the performance of different data augmentation methods with insufficient training dataset. To form new training sub-dataset, we randomly sample partial radio signal samples from the initial 110,000 radio signal training samples, i.e., 12.5\% of the initial training dataset. Then, the LSTM network is trained by feeding the obtained training sub-dataset and is tested with the initial 110,000 radio signal testing samples. Note that 12.5\% of the training dataset is insufficient to train the LSTM network, resulting a low modulation classification accuracy around 45\% under high SNR, as shown in Fig.~\ref{fig9}. After deploying different radio data augmentation methods, the classification accuracy is improved. As expected, both the rotation augmentation and the flip augmentation outperform the Gaussian noise data augmentation. Interestingly, while the training sub-dataset is expanded by a scale factor $N = 4$ after augmentation, in the same size of 50\% of the initial dataset, the rotation/flip augmentation achieves a higher classification accuracy, around 0.04\%-4.03\%, than the baseline by training the LSTM with 50\% of the initial training dataset without augmentation. 

We further consider a joint augmentation policy with both rotation and flip methods, which expands the dataset by a scale factor $N = 6$ (with 2 redundant augmented radio signal samples) as shown in Fig.~\ref{fig4}(a-b). After this joint augmentation, the size of the training dataset is expanded from 12.5\% to be 75\% of the initial training dataset. Interestingly, we obtain similar classification accuracies at different SNRs as the baseline with 100\% training dataset without augmentation, as plotted in Fig.~\ref{fig9}. Note that such a classification accuracy is achieved by using 25\% less training data.

\begin{figure*}[t!] 
	\centering
	\includegraphics[width=7. in]{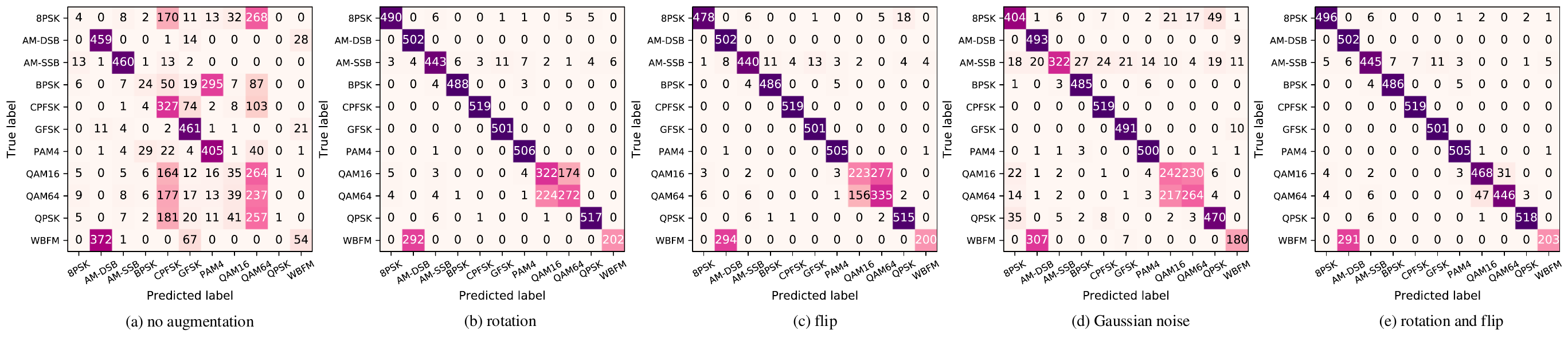}
	\caption{Confusion matrices under different data augmentation methods with 12.5\% training dataset when SNR is 18dB.}\label{fig10}
\end{figure*}

\begin{figure}[t!] 
	\centering
	\includegraphics[width=3.3 in]{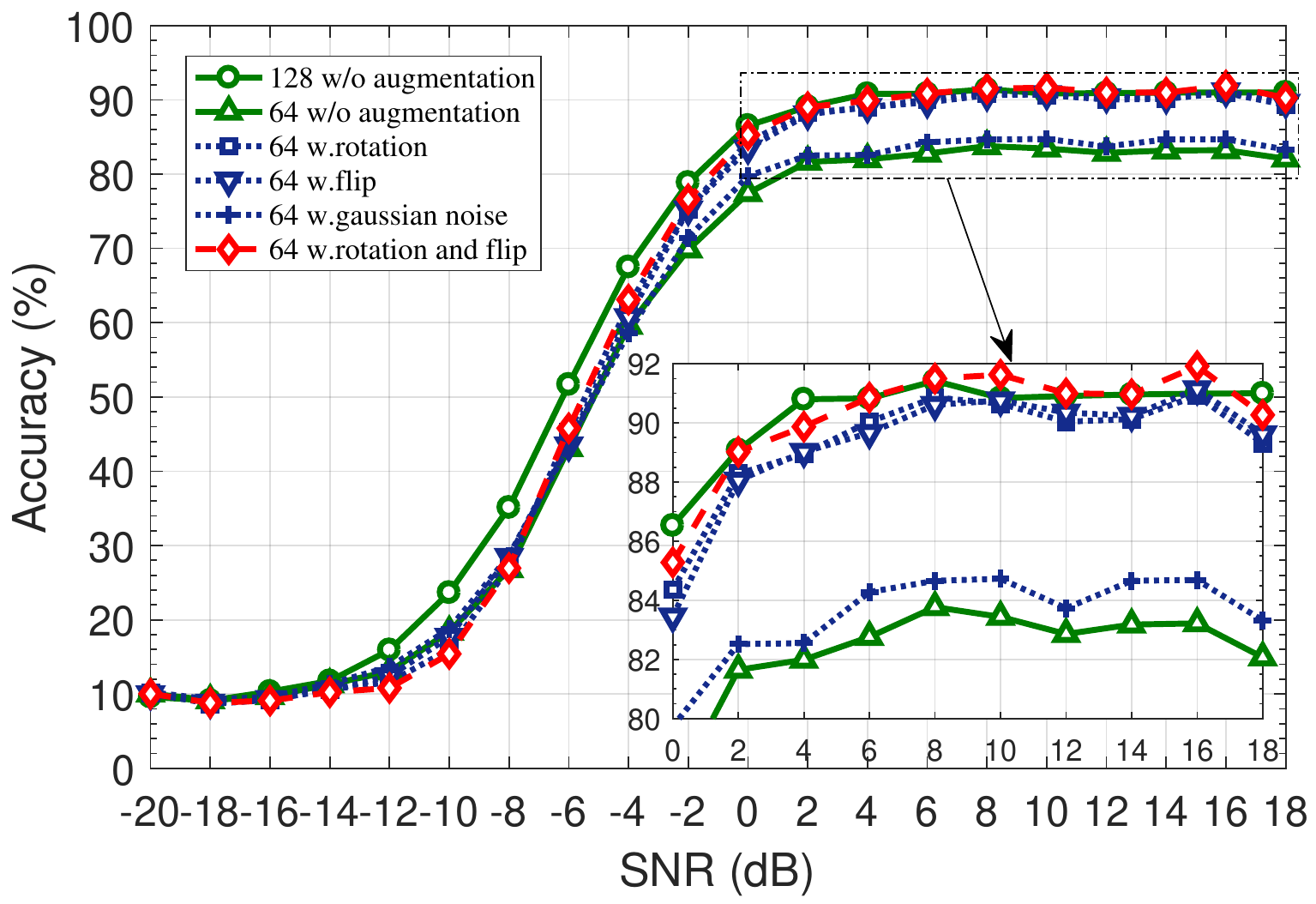}
	\caption{Classification accuracy versus different data augmentation methods under different SNRs with 64-length signal.} \label{fig11}
\end{figure}

To further evaluate the advantages of joint rotation and flip augmentation, we present confusion matrices in different augmentation methods with 12.5\% training dataset at 18dB in Fig.~\ref{fig10}. When training dataset is insufficient, it is difficult to classify BPSK, WBFM, QAM16 and QAM64, whose classification accuracies are significantly improved after joint augmentation. Specifically, in reducing the confusion between QAM16 and QAM64, the joint augmentation performs better than both the rotation augmentation and the flip augmentation.

We have also evaluated another joint augmentation with all three augmentation methods. However, adding Gaussian noise method to the joint rotation and flip augmentation slightly reduces the classification accuracy. Therefore, we conclude that both rotation and flip methods are preferred for radio data augmentation and they can be jointly applied to further improve the augmentation performance.

\subsection{Augmentations On short Sample}
We further evaluate data augmentation methods for modulated radio signals with fewer sampling points. We halve each original 128-point radio signal sample into two new samples and obtained a new dataset consisting of 440,000 entries of 64-point radio signal samples. Similar to previous evaluations, we randomly choose half of them to the LSTM network, which is further tested with the remaining half dataset. With a shorter radio signal sample, the number of LSTM cells in each LSTM layer in Fig.~\ref{fig3} is reduced from 128 to 64, resulting a simpler inference model. Specifically, the number of parameters of the LSTM network is reduced from 201.1K to 54.1K and the inference complexity in FLOPs (floating-point operations) is reduced from 2.8K to 1.4K.

In Fig.~\ref{fig11}, we evaluate modulation classifications with 64-point radio samples. Without augmentation, 64-point modulated radio samples always lead to lower classification accuracy than the baseline with 128-point, around an 8\% reduction when SNR is greater than 0 dB. The classification accuracy is improved after deploying either rotation or flip augmentation. Especially, the joint rotation and flip augmentation can achieve 1\% higher classification accuracy than the baseline under high SNR. Therefore, with  data augmentation, the radio signal modulations can be successfully classified upon receiving only half number of sampling points, which significantly reduces the classification response time.

\section{Conclusion}
In this paper, we studied radio data augmentation methods for deep learning-based modulation classification. Specifically, three typical augmentation methods, i.e., rotation, flip, and Gaussian noise, were studied based on a well-known LSTM model. We first studied radio data augmentations at training and inference phases and revealed that train-test-time augmentation achieves the highest accuracy. Then, we numerically evaluated all three augmentation methods based on the full and partial training dataset. All numerical results show that both the rotation and the flip methods achieve higher classification accuracy than the Gaussian noise method and the rotation method achieves the highest accuracy. Meanwhile, a joint augmentation policy with both rotation and flip methods can further improve the classification accuracy, especially with insufficient training samples. Given only 12.5\% of initial training dataset, the joint augmentation method expands the dataset to be a size of 75\% of the initial dataset and obtains even higher than the baseline with 100\% training datasets without augmentation. Furthermore, after deploying data augmentation, a radio sample can be classified based on only one half of the radio sampling points, resulting in a simplified deep learning model and shorter the classification response time.  

\bibliographystyle{IEEEtran}
\bibliography{IEEEabrv,ref}

\end{document}